\documentstyle[prl,aps]{revtex}

\begin{document}
\title{GRAVITATIONAL WAVE BACKGROUND FROM HYBRID TOPOLOGICAL DEFECTS}
\author{Xavier Martin and Alexander Vilenkin}
\address{Institute of Cosmology, Department of Physics and Astronomy,
Tufts University, Medford, MA 02155, USA}
\maketitle

\begin{abstract}
We investigate the spectrum of stochastic gravitational wave
background generated by hybrid topological defects: domain walls
bounded by strings and monopoles connected by strings.  Such defects
typically decay early in the history of the universe, and their mass
scale is not subject to the constraints imposed by microwave
background and millisecond pulsar observations. Nonetheless, the
intensity of the gravitational wave background from hybrid defects can
be quite high at frequencies above $10^{-8} {\rm Hz}$, and in
particular in the frequency range of LIGO, VIRGO and LISA detectors.
\end{abstract}

The most interesting cosmological events, from the particle physics
point of view, occurred during the first second after the big bang.
However, the universe became transparent to electromagnetic waves only
at the time of decoupling, $t_{dec} \approx 6\times 10^{12} (\Omega
h^2)^{-1/2} s$, and our ability to observe these events is
very limited.  In fact, it appears that direct signals from sources at
$t < 1s$ can arrive to us only in the form of gravitational waves.
A new generation of gravitational-wave detectors will start
operating early next century \cite{Thorne}, and now is an appropriate 
time to think
about the possible signals we can get from the very early universe.

A number of potential sources have already been suggested
\cite{Reviews}: quantum fluctuations of the metric during inflation
\cite{Gw,Brustein}, oscillating loops of cosmic string \cite{V81}, and
colliding bubbles in a first-order phase transition \cite{Turner}.
In this paper we shall investigate the spectrum of gravitational waves
generated by transient topological defects, monopoles connected by
strings and domain walls bounded by strings.

Among the most significant cosmic events are the phase transitions
corresponding to spontaneous symmetry breakings in the early universe.
Such transitions are typically accompanied by the formation of
topological defects \cite{Kibble,Book}.  Depending on the topology of
the symmetry groups involved, the defects can be in the form of
surfaces, lines, or points.  They are called domain walls, strings,
and monopoles, respectively.  Hybrid defects can be formed in a
sequence of phase transitions, e.g., the first transition produces
monopoles, which get connected by strings at the second phase
transition.  Such defects are transient and eventually decay into
relativistic particles.  If this happens at a sufficiently early time,
the decay products thermalize, and we can see no trace of the defects,
except perhaps in the form of gravitational waves \cite{Foot1}. In
the following, we have adopted units such that $c=\hbar =1$.

\paragraph*{Walls bounded by strings.}

The simplest sequence of phase transitions that results in walls
bounded by strings is $G \to H\times Z_2 \to H$.
The first phase transition produces strings, and at the second each
string gets attached to a domain wall.  

The evolution of strings prior to wall formation is
identical to that of ``ordinary'' strings (which do not get attached
to walls).  After a period of overdamped motion, the strings start
moving relativistically and approach a
scaling regime where the characteristic scale of the network is
comparable to the horizon.    
A horizon-size volume at cosmic time $t$ contains a few long
strings stretching across the volume and a large number of small
closed loops.  The typical distance between long strings and their
characteristic curvature radius are both $\sim t$, but in addition the
strings have small-scale wiggles of wavelength down to $l \sim
\alpha t$, with $\alpha \ll 1$.  The typical size of the loops being
chopped off the long strings is comparable to the scale of the
smallest wiggles, $l$.  The loops oscillate and lose their energy by
gravitational radiation.  The lifetime of a loop of size $l$ is $\tau
\sim l/\Gamma G\mu$, where $G$ is the Newton's constant, $\mu$ is the
mass per unit length of string, and $\Gamma \sim 50$ is a numerical
coefficient.   

The exact value of the parameter $\alpha$ is not known.  Numerical
simulations of string evolution \cite{Bennett} suggest that $\alpha
\lesssim 10^{-3}$, while the analysis of gravitational radiation
back-reaction indicates that $\alpha \gtrsim \Gamma G\mu$.  Thus, we
expect $\alpha$ to be in the range $10^{-3}>\alpha >\Gamma G\mu$ for
$G\mu \ll 10^{-5}$ and $\alpha \sim \Gamma G\mu$ for $G\mu \gtrsim
10^{-5}$.  We note that for $\alpha \sim \Gamma G\mu$ the loops decay
in about one Hubble time from their formation.

The evolution of the network after domain walls form, having the
strings as their boundaries, depends on the relative magnitude of the
wall tension $\sigma$, and the force per unit length on a curved
string due to the string tension $\mu$, $F\sim\mu/t$, where we have
assumed that the curvature radius of string is $R\sim t$. If the
walls are formed at $t_w \ll \mu/\sigma$, then initially they have
little effect on string dynamics.  The walls become dynamically
important at $t\sim \mu/\sigma$, when they pull the strings towards
one another, and the network breaks into pieces of wall bounded by
string.  Alternatively, if the walls form at $t_w \gtrsim \mu/\sigma$,
then the breakup of the network occurs immediately after the wall
formation.  We shall use the notation $t_s$ for the time when the
strings start moving relativistically and $t_* = max \{\mu /\sigma,
t_w\}$ for the time of the network breakup.  The typical size of the
pieces is expected to be $\sim \alpha t_*$.

Gravitational waves emitted by oscillating loops of string during the
period $t_s < t < t_*$ add up to a stochastic gravitational-wave
background with a nearly flat spectrum \cite{V81,VV85,BB90,Caldwell},
\begin{equation}
\Omega_g (\nu) \equiv {\nu\over{\rho_c}}{d\rho_g\over{d\nu}}\sim 300
\left( {\alpha G\mu \over{\Gamma}}\right)^{1/2} \left({{\cal N}\over
{\cal N}_\nu}\right)^{1/3}\Omega_r.
\label{spectrum}
\end{equation}
Here, $\rho_g$ is the energy density of gravitational waves, $\rho_c$
is the critical density, $\nu$ is the frequency, 
${\cal N}(t)$ is the effective number of spin degrees
of freedom in the cosmic plasma at time $t$, ${\cal N}_\nu$ is the
value of ${\cal N}$ when waves of frequency $\nu$ were emitted, and
$\Omega_r = \rho_r/\rho_c$, where $\rho_r$ is the energy density in
radiation and relativistic particles.  The spectrum (\ref{spectrum})
extends over a range of frequencies, $\nu_{min}<\nu<\nu_{max}$, which
is determined by the times $t_s$ and $t_*$.  For a rough estimate, we
can use a simple model which assumes that the loops emit most of their
energy in the first few harmonics with frequencies $\nu\sim 2/l$,
where $l$ is the loop's length.  To simplify the discussion, we shall
assume also that $t_*$ is in the radiation era and that 
$\alpha\sim\Gamma G\mu$ (which should be reasonably
accurate for superheavy strings with $G\mu \gtrsim 10^{-6}$).  Then,
gravitational waves of present frequency $\nu$ were emitted at
\begin{equation}
t_\nu \sim 10^{-10}\mu_{-6}^{-2} \left( {\nu\over{1 {\rm Hz}}}
\right)^{-2} s,
\label{tnu}
\end{equation}
where $\mu_{-6}=G\mu /10^{-6}$ and we have used the approximate form,
$a \propto t^{1/2}$, for the scale factor during the radiation era.
The cutoff frequencies $\nu_{min}$ and $\nu_{max}$ correspond to the
times $t_*$ and $t_s$, respectively.  In a more realistic model, with
a full account taken of the radiation spectrum by the loops, the
cutoffs would be spread over about two orders of magnitude in
frequency, while the middle part of the spectrum (\ref{spectrum})
would be unaffected.

According to Eq.(\ref{tnu}), gravitational waves in the frequency
range of the LIGO and VIRGO detectors ($\sim 100$ Hz) were emitted at
$t_{LV} \sim 10^{-14}\mu_{-6}^{-2} s$, while waves in the sensitivity
range of the millisecond pulsar ($\sim 10^{-8}$ Hz) \cite{T}
were emitted at
$t_{MP} \sim 10^6 \mu_{-6}^{-2} s$.  For $t_{LV}<t_*<t_{MP}$, the
gravitational wave background is in the frequency range accessible to
LIGO/VIRGO, but not to the pulsar.  Since in this case the strings
decay before decoupling, they leave no trace on the microwave
radiation, and thus the only constraint on the possible values of
$G\mu$ comes from the nucleosynthesis considerations \cite{Davis}.

At the time of nucleosynthesis, the total energy density in
gravitational waves, in terms of the critical density, is given by
\begin{equation}
\Omega_g ({\rm nucl})\sim 150 G\mu \left({{\cal N}_{nucl}\over{{\cal
N}_*}}\right)^{1/3} \ln \left({t_* \over{t_s}}\right).
\label{nucl}
\end{equation}
Here, ${\cal N}_*$ and ${\cal N}_{nucl}\sim 10$ are, respectively, the
values of ${\cal N}$ at $t_*$ and at the time of nucleosynthesis,
$t_{nucl}\sim 1$ s, and we have assumed for simplicity that the value
of ${\cal N}$ does not change through the whole period from $t_s$ to
$t_*$.  For the standard nucleosynthesis scenario to work, $\Omega_g
({\rm nucl})$ should not exceed 0.054.  With ${\cal N}_*\sim 100$, the
resulting bound on $G\mu$ is
\begin{equation}
G\mu \lesssim \frac{8\times 10^{-4}}{\ln (t_*/t_s)}.
\label{nucbound}
\end{equation}
With the same assumptions, Eq. (\ref{spectrum}) gives
\begin{equation}
\Omega_g (\nu)\sim 6\times 10^{-3}G\mu h^{-2}.
\label{s}
\end {equation}

The logarithm in Eq.(\ref{nucbound}) can take values from $\sim 2$ to
$\sim 100$.  Consider, for example, strings with $G\mu\sim 10^{-5}$
and walls with symmetry breaking scale $\eta_w \sim 100$ GeV.  If the
symmetry breaking potential for the walls does not have very small
couplings, then $\sigma\sim \eta_w^3$ and $t_*\sim 1000$ s.  The time
$t_s$ when the strings start moving relativistically  depends on the
cosmological scenario.  In some models this will be the time when the
damping force due to the interaction of strings with plasma becomes
smaller than the force of tension in the strings, $t_s\sim (G\mu)^{-2}
t_p \sim 10^{-33} s$, where $t_p$ is the Planck time.  Then, $\ln
(t_*/t_s) \approx 80$, and the constraint (\ref{nucbound}) is
marginally satisfied.

The time $t_s$ can be greatly increased in inflationary models where
the string-forming phase transition occurs during inflation, but
sufficiently close to its end, so that the strings are not completely
inflated away.  In this case, $t_s$ is the time when the typical
distance between the strings becomes smaller than the Hubble radius.
Strings can also be produced in a ``pre-heating'' transition after
inflation \cite{KLS}.  The amplitude of scalar field fluctuations at
pre-heating can reach Planckian values, and superheavy defects can be
formed even if the thermalization temperature after inflation is very
low.  The time $t_s$ is then shortly after the end of inflation.  In
some supersymmetric models, certain couplings can be naturally small,
and superheavy strings can be formed as late as the electroweak phase
transition \cite{LSP}.  Then, $t_s\sim t_{ew}\sim 10^{-11} s$.
Finally, the effective value of $t_s$ can be increased by massive
particle annihilations (that is, by a decrease of ${\cal N}$ between
$t_s$ and $t_*$), resulting in a dilution of gravitational waves from
earlier epochs.  

The time $t_*$ is, of course, also model-dependent (through the
parameters $\mu$ and $\sigma$), and in general $t_s$ and $t_*$ do not
have to be separated by many orders of magnitude.  With $\ln
(t_*/t_s)\sim 1$, the constraint (\ref{nucbound}) requires only that
$G\mu \lesssim 8\times 10^{-4}$, and Eq.(\ref{s}) gives $\Omega_g
(\nu) \lesssim 5\times 10^{-6} h^{-2}$. The first version of LIGO
interferometer will be able to detect a stochastic background with
$\Omega_g (\nu) \gtrsim 2\times 10^{-6}h^{-2}$ after two years of
observations, at frequencies $\sim 100$ Hz and with a $90\% $ confidence
\cite{Reviews}, and thus a detection of gravitational waves from a
string-wall network is only marginally possible. Advanced LIGO and
VIRGO will be sensitive to $\Omega_g (\nu ) \gtrsim 5\times 10^{-11}
h^{-2}$, which corresponds to $G\mu \gtrsim \times 10^{-8}$. A
similar sensitivity will be achieved by LISA space-based
interferometer at $\nu \sim 10^{-3}Hz$.

\paragraph*{Monopoles connected by strings.}

The prototypical sequence of symmetry breakings resulting in monopoles
connected by strings is $G \to H\times U(1) \to H$.  Monopoles formed
at the first phase transition get connected by strings at the second
phase transition.  If the monopole-forming phase transition occurs
after inflation, then the average monopole separation is always
smaller than the Hubble radius, and when monopole-antimonopole
($M{\bar M}$) pairs get connected by strings and start oscillating,
they typically dissipate the bulk of their energy to friction in less
than a Hubble time \cite{Book,HKR}.  This does not result in any
appreciable gravitational-wave background.

The most interesting scenario, for our purposes, is when monopoles are
formed during inflation (but are not completely inflated away).  
Strings can either be formed later during inflation,
or in the post-inflationary epoch.  The characteristic length scale of
strings at formation is then much smaller than the monopole
separation; the strings connecting monopoles have Brownian shapes, and
there is also a large number of closed loops.  The evolution of
strings is initially identical to that of topologically stable
strings, without monopoles.  At $t\sim t_s$, the strings start moving
relativistically and generate a gravitational-wave background of
intensity (\ref{spectrum}).  In the course of the evolution, the
characteristic length of strings grows like $t$ and eventually becomes
comparable to the monopole separation, so that we are left with
$M{\bar M}$ pairs connected by more or less straight strings.  We
shall call the time when this happens $t_m$ and the monopole
separation at that time, $d$ (note that $d\sim t_m$).  At $t>t_m$,
$M{\bar M}$ pairs oscillate and gradually lose their energy by
gravitational radiation.

In the above discussion we have assumed implicitely that $t_m > t_s$.
In this case, using the inequalities $d\sim t_m >t_s \gtrsim
(G\mu)^{-2} t_p$, it is easily verified that
\begin{equation}
\mu d \gg m,
\end{equation}
where $m$ is the monopole mass.  Then the motion of monopoles is
relativistic, with a typical Lorentz factor
\begin{equation}
\gamma\sim\mu d/m \gg 1.
\label{gamma}
\end{equation}
If strings are formed during inflation, soon after the monopoles, then
the strings connecting $M{\bar M}$ pairs can be nearly straight.  In
this case, the strings do not go through a period of relativistic
evolution, and no gravitational radiation is produced prior to $t_m$.
But unless these defects are formed very close to the end of
inflation, the inequality (\ref{gamma}) will still be satisfied.  We
shall assume this to be the case.

Oscillating $M{\bar M}$ pairs lose their energy by gravitational
radiation and by friction due to scattering of plasma particles off
the monopoles \cite{Foot2}.  To estimate the rate of gravitational
energy loss, we calculated the power and the spectrum of gravitational
waves radiated by an oscillating pair connected by a straight string.  
Details of this calculation will be given elsewhere \cite{MV}, while
here we 
shall only summarize the relevant features of the radiation spectrum.
Most of the energy of an oscillating $M{\bar M}$ pair is radiated in
the frequency range
\begin{equation}
d^{-1} \lesssim \nu \lesssim \gamma^2 d^{-1}
\label{range}
\end{equation}
with a spectrum
\begin{equation}
d{\cal E}_g/dt d\nu\approx 4G\mu^2\nu^{-1}.
\label{sp}
\end{equation}
At higher frequencies, $d{\cal E}_g/dt d\nu \propto \nu^{-2}$.
The total radiation power is \cite{Foot3} 
\begin{equation}
{\dot {\cal E}_g}=\tilde{\Gamma} G\mu^2,
\label{edot}
\end{equation} 
where 
\begin{equation}
\tilde{\Gamma} \approx 8\ln\gamma \label{gamtild}
\end{equation}
is a numerical coefficient taking values in the range $10
\lesssim \tilde{\Gamma} \lesssim 10^3$, depending on the Lorentz 
factor $\gamma$.  

A monopole-antimonopole pair connected by a straight
string is, of course, a very special configuration, and one could be
concerned that the radiation spectrum in this case may be very
different from that for a generic configuration.  To address this
issue, we calculated the spectrum of electromagnetic radiation emitted
by a pair of oscillating, equal and opposite charges connected by a
straight string.  We found that the resulting spectrum is
qualitatively similar to that in the generic case \cite{Schwinger}.
This suggests that the same may be true in the case of gravitational
radiation, but of course a detailed analysis is required before any
definite conclusion can be drawn.  Here, we shall assume that the
spectrum (\ref{sp}) is representative of the generic case \cite{butcom}.

The frictional energy loss is of the order 
\begin{equation}
{\dot{\cal E}_f}\sim T^2,
\end{equation}
where $T$ is the temperature, and it is easily verified that
${\dot{\cal E}_f}\ll{\dot{\cal E}_g}$ for $t_m\gg t_s$.  Hence, the
lifetime of a pair is
\begin{equation}
\tau\sim\mu d/{\dot{\cal E}_g}\sim d/\tilde{\Gamma} G\mu.
\end{equation}

The number density of $M{\bar M}$ pairs at $t\sim t_m\sim d$ is $n_M
(t_m) \sim d^{-3}$, and at later times $n_M (t)\sim (td)^{-3/2}$
$\quad (t_m < t <\tau)$.  The spectral power of the gravitational wave
background 
emitted by the oscillating pairs per Hubble time at time $t$ is
\begin{equation}
\Omega_g^{(t)}(\nu)\sim 4G\mu^2 t n_M(t)/\rho_c(t) \sim 120 (G\mu)^2
(t/d)^{3/2},
\end{equation}
where $\rho_c(t)\sim 1/30Gt^2$ is the critical density.   The highest
power comes from the radiation emitted at $t\sim \tau$.  The frequency
range (\ref{range}) at that time corresponds to the present range
\begin{equation}
\nu_* \lesssim \nu \lesssim \gamma^2\nu_*,
\label{newrange}
\end{equation}
where
\begin{equation}
\nu_*\sim d^{-1} (\tau/t_{eq})^{1/2}z_{eq}^{-1},
\label{nustar}
\end{equation}
$t_{eq}$ is the time of equal matter and radiation densities, and
$z_{eq}$ is the corresponding redshift.  The spectral power in the
range (\ref{newrange}) is [compare with Eq.(\ref{spectrum})]
\begin{equation}
\Omega_g (\nu)\sim 120\tilde{\Gamma}^{-3/2}(G\mu)^{1/2}({\cal N}/{\cal
N}_\tau)^{1/3} \Omega_r .
\label{newspectrum}
\end{equation}

For $\nu<\nu_*$, the main contribution to the spectrum is made at the
time $t_\nu\sim (\nu/\nu_*)^2\tau$, when the frequency corresponding
to $\nu$ at the present time is $\sim d^{-1}$.  The resulting spectral
power is given by Eq.(\ref{newspectrum}) with an additional factor
$(\nu /\nu_*)^3$.  This part of the spectrum extends to the minimum
frequency $\nu_{min}\sim (\tilde{\Gamma} G\mu)^{1/2}\nu_*$.

Radiation emitted prior to $t_{nucl}\sim 1s$ should satisfy the
nucleosynthesis constraint.  Assuming that $\tau < t_{nucl}$, the
total energy density in gravitational waves
at $t_{nucl}$ is
\begin{equation}
\Omega_g(nucl)\sim 30 (G\mu/\tilde{\Gamma})^{1/2}({\cal N}_{nucl} / 
{\cal N}_\tau)^{1/3}\lesssim 0.05,
\label{newnuc}
\end{equation}
where we have used Eq.(\ref{gamtild}). Combining this with 
(\ref{newspectrum}), we have
\begin{equation}
\Omega_g(\nu) \lesssim 10^{-5}\tilde{\Gamma}^{-1}h^{-2}
\end{equation}

For the frequency range (\ref{newrange}) to overlap with the
sensitivity range of LIGO/VIRGO, it is necessary that $d \gtrsim 
2\times 10^{-13} (\tilde{\Gamma} G\mu)^{-1} cm$, which implies
$\tilde{\Gamma} \lesssim 400$ (assuming that $m<m_p$, where
$m_p$ is the Planck mass), and thus $\Omega_g (\nu)
\lesssim 2\times 10^{-8}h^{-2}$.  This is below the sensitivity of the
first version of LIGO but may be within the reach of the advanced
detectors.

If the strings did have a period of relativistic evolution, then the
spectrum includes another flat region with spectral power (\ref{s}) at
$\nu > (\tilde{\Gamma} G\mu)^{-1/2}\nu_*$.  This power is greater than
(\ref{newspectrum}) if $G\mu >\tilde{\Gamma}^{-3}$ which, for
$\tilde{\Gamma} \sim 10^3$, includes all interesting values of $G\mu$.
The analysis of this part of the spectrum is identical to that for
walls bounded by strings, with $t_*$ replaced by $t_m$.  As before,
the advanced detectors will have sufficient sensitivity in the case
of strings with $G\mu \lesssim \times 10^{-8}$, provided that
the flat portion of the spectrum includes $\nu\sim 100 Hz$.

This work was supported in part by the National Science Foundation.


\begin{thebibliography}{11}
\bibitem{Foot1}There are some notable exceptions to this rule.  If
the constituent fields of the defects have baryon-number-nonconserving
couplings, then the decay of transient defects can result in a net
baryon asymmetry of the universe.  Also, in the case of global
defects, a background of very weakly interacting goldstone or
pseudo-goldstone bosons can be generated.
\bibitem{Foot2}Here we assume that monopoles do not have unconfined
magnetic charges (that is, the only magnetic charge they have is the
one whose flux is squeezed into the strings).  Otherwise, monopoles
would also lose energy by radiating gauge quanta.
\bibitem{Foot3}Note that Eq.(\ref{edot}) could be obtained as a rough
estimate using the quadrupole formula, ${\dot {\cal E}_g}\sim G{\cal
E}^2 d^4\nu^6\sim G\mu^2$, where ${\cal E}=\gamma m\sim \mu d$ is the
monopole energy and $\nu\sim d^{-1}$ is the characteristic frequency
of oscillation.  However, the validity of the quadrupole formula for
ultrarelativistic monopoles is rather dubious.
\bibitem{butcom}At large frequencies, the $\nu^{-2}$ behaviour comes
{}from the discontinuity in the monopoles acceleration when they pass
one another and the string tension is reversed. In a more general
case, the acceleration should vary smoothly and the spectrum will fall
exponentially.
\end{thebibliography}
\end{document}